# Prenatal stress perturbs fetal iron homeostasis in a sex specific manner

Peter Zimmermann[1]; Marta C. Antonelli[1,2]; Ritika Sharma[1,3]; Alexander Müller[4]; Camilla Zelgert[1]; Bibiana Fabre[5]; Natasha Wenzel[6]; Hau-Tieng Wu[7,8]; Martin G. Frasch[9*¶]; Silvia M. Lobmaier[1*¶].

[1]Department of Obstetrics and Gynecology, Klinikum rechts der Isar, Technical University of Munich, Germany; [2]Instituto de Biología Celular y Neurociencias "Prof. E. De Robertis," Facultad de Medicina, UBA, Buenos Aires, Argentina; [3]Helmholtz Zentrum Munich, Neuherberg, Germany; [4]Innere Medizin I, Department of Cardiology, Klinikum rechts der Isar, Technical University of Munich, Germany; [5]Facultad de Farmacia y Bioquímica, Instituto de Fisiopatología Y Bioquímica Clínica (INFIBIOC), Universidad de Buenos Aires, Buenos Aires, Argentina; [6]Department of Epidemiology, University of Washington, Seattle, WA, USA; [7]Dept. of Mathematics and Dept. of Statistical Science, Duke University, Durham, NC, USA; [8]Mathematics Division, National Center for Theoretical Sciences, Taipei, Taiwan; [9]Department of Obstetrics and Gynecology and Center on Human Development and Disability (CHDD), Research Unit of Molecular Epidemiology, Institute of Epidemiology, University of Washington, Seattle, WA, USA.

* corresponding author; E-mail: mfrasch@uw.edu

¶*these authors contributed equally to this work*

# Abstract

The adverse effects of maternal prenatal stress (PS) on child's neurodevelopment warrant the establishment of biomarkers that enable early interventional therapeutic strategies.

We performed a prospective matched double cohort study screening 2000 pregnant women in third trimester with Cohen Perceived Stress Scale-10 (PSS-10) questionnaire; 164 participants were recruited and classified as stressed and control group (SG, CG). Fetal cord blood iron parameters were measured at birth. Transabdominal electrocardiograms-based Fetal Stress Index (FSI) was derived. We investigated sex contribution to group differences and conducted causal inference analyses to assess the total effect of PS exposure on iron homeostasis using a directed acyclic graph (DAG) approach. Differences are reported for $p<0.05$ unless noted otherwise. Transferrin saturation was lower in male stressed neonates. The minimum adjustment set of the DAG to estimate the total effect of PS exposure on fetal ferritin iron biomarkers consisted of maternal age and socioeconomic status: SG revealed a 15% decrease in fetal ferritin compared with CG. Mean FSI was higher among SG than among CG. FSI-based timely detection of fetuses affected by PS can support early individualized iron supplementation and neurodevelopmental follow-up to prevent long-term sequelae due to PS-exacerbated impairment of the iron homeostasis.

# Introduction

In the second and third trimester of pregnancy maternal iron requirements can increase up to eightfold or a total of 1 g of additional iron, due to expanding maternal and fetal erythropoiesis[1,2]. Iron homeostasis dysregulation of pregnant mothers and/or children is known to induce lasting neurological damage in the offspring[3].

Prenatal maternal stress (PS), including both pregnancy-specific and general psychosocial stress and anxiety, can jeopardize the balance of the maternal iron homeostasis[4-7]. Pregnant women are especially vulnerable to chronic stress as they face new and potentially challenging situations such as body image issues, lifestyle changes, and fluctuating hormones[8]. PS induces lasting changes to fetal stress response, in a process known as "fetal programming"[9] which might be partly transmitted by the autonomic nervous system (ANS) and the hypothalamic–pituitary–adrenal (HPA) system. In an interim analysis of pregnant women with PS and controls, we showed that PS results in entrainment of fetal heart rate (fHR) by maternal heart rate (mHR), thus yielding a non-invasively obtainable PS biomarker in mother–fetus dyads that we refer to as Fetal Stress Index (FSI)[10]. HPA dysregulation increases the risk of newborn impairment and higher vulnerability toward certain chronic diseases and neurobehavioral disorders[11,12].

Sex-specific PS effects are well described and recommended for the general consideration as part of human PS studies[13]. However, the contribution of child's sex to the PS-induced alterations in iron homeostasis of the neonate is unclear and suffers from contradictions[4].

Consequently, we tested the hypothesis that PS influences the fetal ANS which results in sex-specific changes to FSI during third trimester and the iron homeostasis in human neonates.

To gauge the fetal iron homeostasis, we assessed the following iron biomarkers in umbilical cord blood serum: ferritin, transferrin, hepcidin[14] and iron[1]. The putative relationships between hepcidin, PS and ANS are summarized in Fig 1.

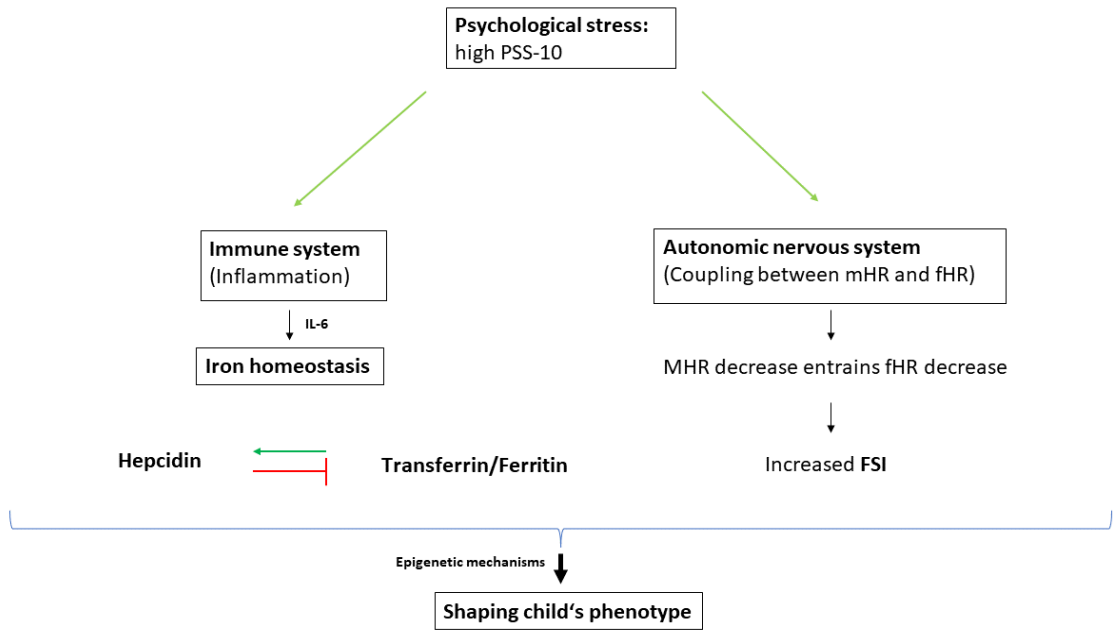

**Figure 1. Putative link of prenatal maternal stress (PS) with iron homeostasis and maternal–fetal heart rate coupling.**

# Results

## Sociodemographic parameters and perinatal outcomes

Women enrolled in the study had a mean age of 33.0 (±4.4) years and a median gestational age of 36.4 (35.3–37.4) weeks of gestation at study entry. Demographics of the excluded participants did not differ from those with all measures. Statistical comparison of SG and CG showed no differences in the used matching criteria (Table S1).

## Maternal and fetal iron homeostasis

10.4% of the included women were anemic prior to delivery (Hb < 11mg/dL). However, we found no differences in fetal iron parameters, maternal intake of iron supplements, fetal and maternal hemoglobin, RBC indices and anemia status between SG and CG (Table S2).

## FSI

MHR and fHR coupling analysis revealed a higher FSI among SG than among CG (0.38 ((–0.22)–0.75) versus –0.01 ((–0.36)–0.34); p = 0.024) (Table S1). FSI showed no correlation to any measured fetal iron biomarker for either sex (Table S3).

However, within the automatically binned ranges of cord blood serum iron biomarker values, we observed FSI differences. FSI was higher in SG than in CG for ferritin levels between 153 and 279 µg/L ((n=42; 24 CG; 18 SG); 0.40 (±0.57) versus 0.01 (±0.47); p = 0.03, transferrin saturation of 32%–47% ((n=24; 12 CG; 12 SG); 0.30 (±0.66) versus –0.24 (±0.27); p = 0.045), and hepcidin values between 0 and 57 ng/mL ((n=92; 47 CG; 45 SG) (0.34 (±0.68) versus –0.01 (±0.56); p = 0.01). Using current newborn guidelines and validated ranges the above-mentioned values of iron markers would be normal[15-17]. Overall, FSI at ~36 weeks of gestation was higher in SG fetuses averaging 0.34 compared with –0.10 in CG fetuses within these cord blood iron biomarker ranges.

## Sex-specific differences

We identified sex-specific differences in iron homeostasis among male infants and showed that the PS effect on iron homeostasis depends on the neonates' sex.

Cord blood transferrin saturation was lower in SG male neonates compared with those in male CG, regardless of iron supplementation. For ferritin levels, we observed a trend towards lower values in male SG (Table 1).

The GEE model revealed that sex is a significant effect modifier that exhibited differences for ferritin ($p = 0.038$, Fig 2), and a trend for transferrin saturation ($p = 0.070$, Fig S1). For hepcidin, we found no significant sex-driven differences.

Interestingly, maternal hair cortisol tended to increase in SG mothers of female neonates (Table 1). FSI group differences were explained by male neonates only.

**Table 1. Sex-specific effect of PS on biomarkers.**

| Characteristics | CG | SG | |
|---|---|---|---|
| Male newborns | n=26 | n=32 | p |
| FSI (n=35 CG, n=43 SG) | –0.13 ((–0.45)–0.31) | 0.30 ((–0.18)–0.61) | **0.050** |
| Maternal hair cortisol [pg/mg] (n=35 CG, n=36 SG) | 115 (14–146) | 124 (40–161) | 0.466 |
| Cord blood ferritin [µg/L]* | 229.7 (113.9–429.6) | 149.6 (96.8–234.0) | **0.069** |
| Cord blood transferrin saturation [%] | 63.4 (±17.7) | 52.9 (±20.2) | **0.041** |
| Cord blood hepcidin [ng/dL] | 26.1 (11.8–41.8) | 17.0 (10.5–30.7) | 0.184 |
| Female newborns | n=28 | n=21 | p |
| FSI (n=39 CG, n=22 SG) | 0.10 (±0.55) | 0.27 (±0.84) | 0.394 |
| Maternal hair cortisol [pg/mg] (n=32 CG, n=21 SG) | 88 (46–119) | 122 (67–180) | **0.073** |
| Cord blood ferritin [µg/L] | 218.2 (±84.8) | 243.1 (±130.0) | 0.423 |
| Cord blood transferrin saturation [%] | 55.9 (±17.0) | 57.8 (±17.8) | 0.703 |
| Cord blood hepcidin [ng/dL] | 22.7 (14.1–37.7) | 20.9 (6.0–37.1) | 0.599 |

Data are mean (SD) using t-test or median (interquartile range) using Mann-Whitney U test. Sample size is indicated as applicable. Differences with p-value < 0.1 are in bold.

*missing values for 1 SG

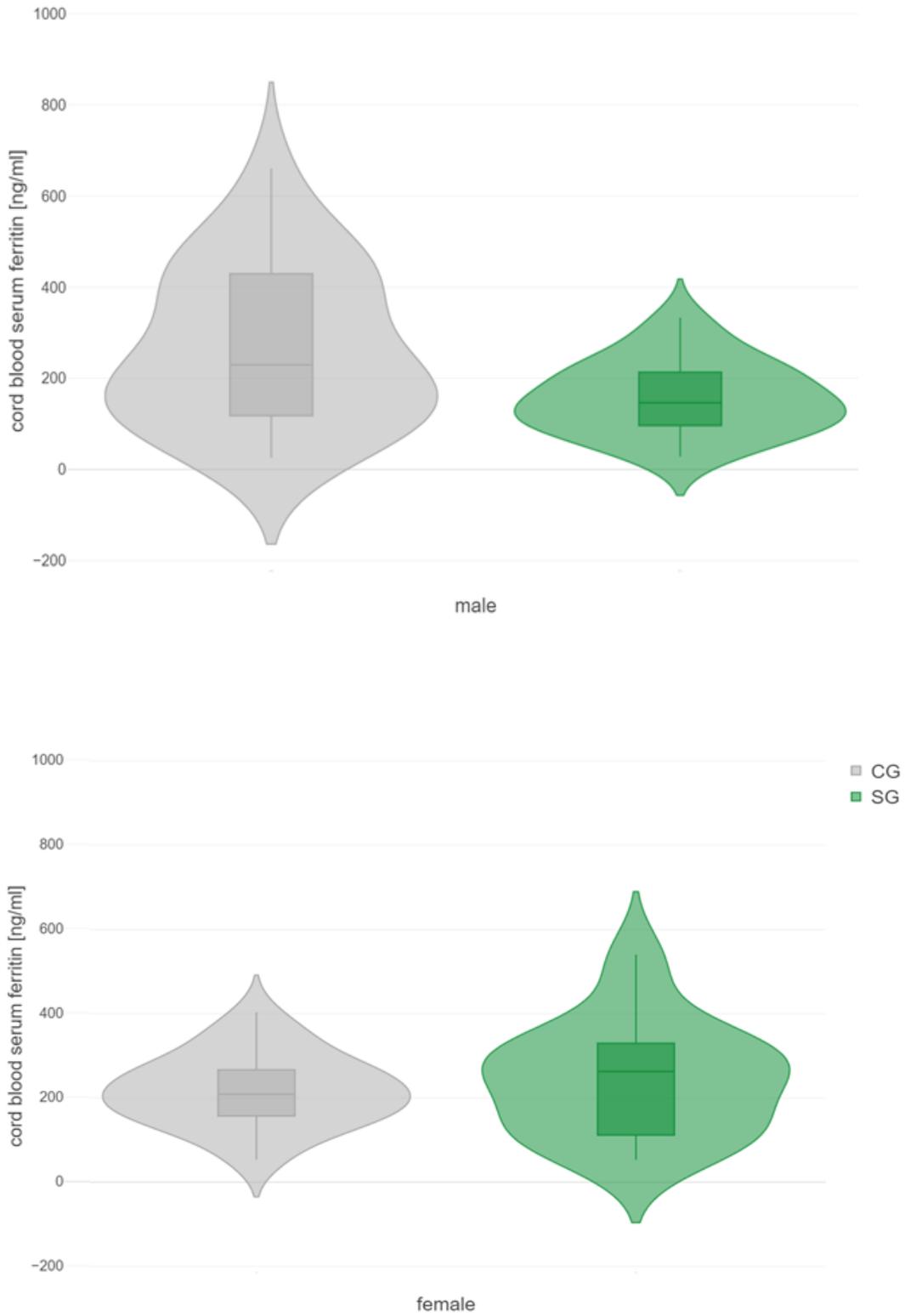

**Figure 2. Sex-dependent group difference in cord blood serum ferritin levels.**

## Estimated causal effect

We conducted causal inference to investigate the effects of the aforementioned relationships more explicitly. We identified certain variables as adjustment sets in blocking all non-causal paths between the treatment and outcome variables while leaving all causal paths unblocked (Fig. 3). Examination of the causal model on the *PS → Cord Blood Ferritin* and *PS → Bayley Score* pathway demonstrated two minimum adjustment sets: "Maternal Age" and "SES" or "Maternal Age" and "Education." Either set could be used to obtain an estimate for the causal effect. Our SES data are represented by "Household income>5000€/month," and maternal education by "University Degree." Controlling for the minimum adjustment set "University Degree" and "Maternal Age" revealed an estimated average exposure effect of lowered cord blood ferritin at the alpha = 0.10 level at –38.06 μg/L (95% CI: –79.91 to 3.78) in SG compared with that in CG (Table S4). This average exposure effect became obscured when fetal sex was included (p-value increased from 0.07 to 0.19) demonstrating that sex is a strong effect modifier on the causal pathway between *PS → Fetal Iron Biomarker*.

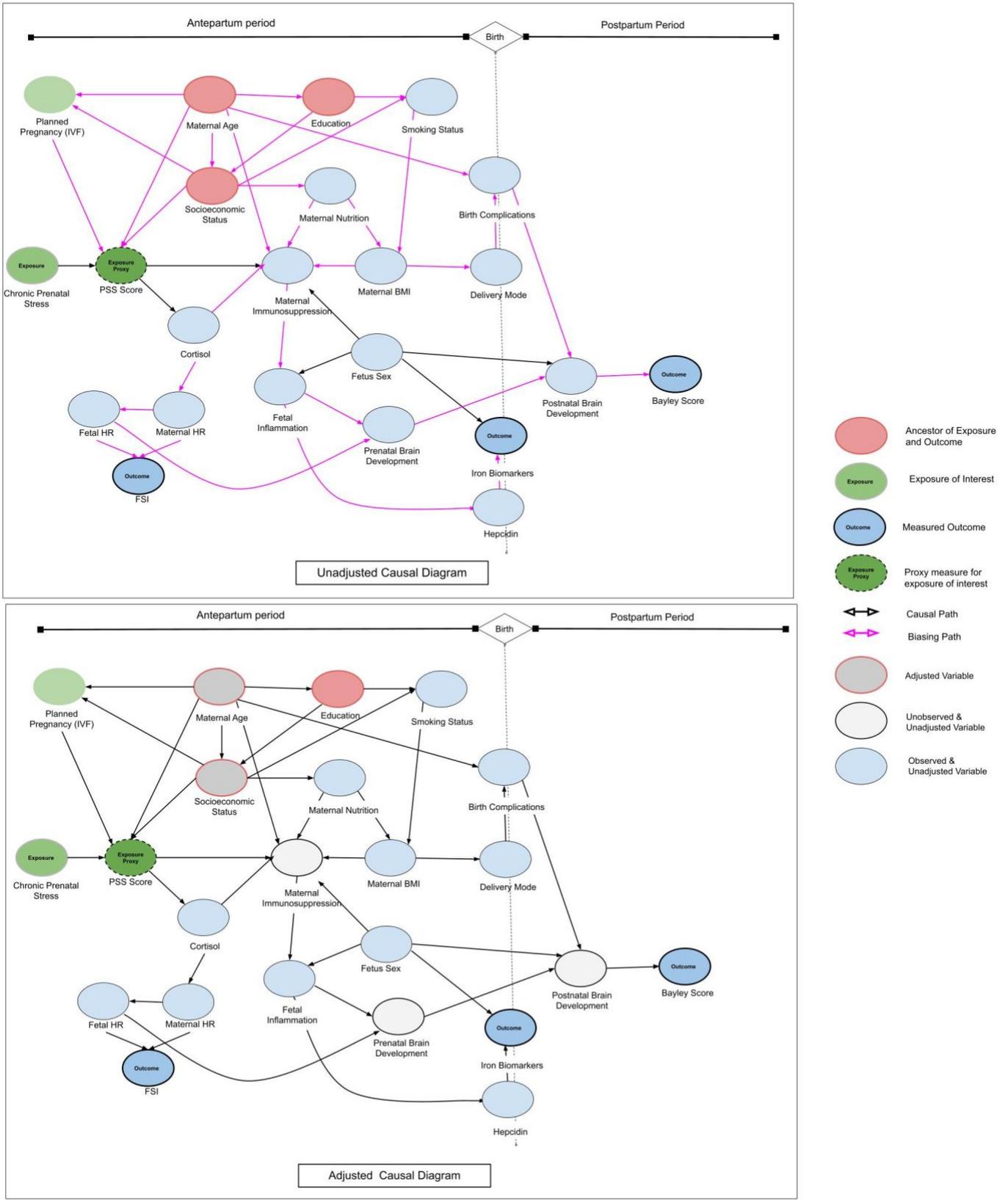

**Figure 3. Maternal age, socioeconomic status, and education as confounding factors within the prenatal stress trajectory.**

## ML for group classification

First, we considered iron biomarkers and FSI, which predicted the groups at an AUROC = 0.706 (±0.194). Next, we added the salient clinical and demographic data (gestational and maternal age, BMI at study entry, pre-pregnancy BMI, planned/non-planned pregnancy, higher education yes/no, income over 5000€ yes/no). These are the features that we also used in the DAG approach and that were available at the time of the taECG measurement at study entry. Doing so, we achieved an AUROC=0.759 (±0.082). The corresponding feature importance ranking is shown in the supplement (Fig. S2). Use of clinical and demographic data reduced the classification performance to AUROC = 0.688 (±0.142), similar to using FSI (AUROC = 0.665 ± 0.126) or iron parameters (AUROC = 0.587 ± 0.183) alone. In general, sex ranked in the lower 5% of variable importance, yielding only a slight improvement.

# Discussion

## PS disrupts fetal iron homeostasis in a sex-specific manner

This study indicates a sex-dependent difference in fetal iron homeostasis and FSI due to PS in an otherwise healthy cohort, mainly driven by the male sex. Causal inference approach allowed us to independently verify fetal sex as an important effect modifier on the causal pathway between PS and cord blood ferritin. The findings strengthen previous published FSI results[10].

The PS effect on the fetal iron biomarkers has been poorly understood. Rhesus monkey infants born to stressed mothers were more likely to develop iron deficiency[7]. Likewise, several studies in humans have shown a correlation between PS and cord blood zinc protoporphyrin/heme index as well as PS and ferritin levels[4-6].

During pregnancy, maternal stress hormones such as cortisol influence the growing fetus and its neurodevelopment, presumably via epigenetic mechanisms[9,18]. Cortes and colleagues proposed an influence of chronic stress through a stress-induced altered expression of a variant of the enzyme acetylcholinesterase on the iron-regulating system in fetal sheep brain-derived primary microglia cultures[19]. They assumed the afferent cholinergic anti-inflammatory pathway signaling on microglial α7 nicotinic acetylcholine receptors to down-regulate metal ion transporter and ferroportin, which acts as a hepcidin receptor (Fig 1).

Animal studies observed stress-dependent cognitive deficits mainly seen in males[20,21]. In humans, sex-specific PS effects are reflected by lower scores in conduct assessments and higher test scores for emotional disturbance in males compared to females[13,22,23]. Campbell et al. applied six specific PS questionnaires, each twice in the second and third trimester, to 428 ~28-years-old mothers and found newborns of pregnant women exposed to violence to be stronger associated with cord blood ferritin levels lower in boys than in girls[4].

## The relation of iron homeostasis biomarkers to PS

Our results show no relationship between the presence of maternal anemia, fetal iron deficiency and PS. These findings are in agreement with literature suggesting that the fetus is robust against moderate changes of the maternal iron homeostasis[24,25].

Within the DAG framework, we estimated that PS reduced the cord blood serum ferritin levels by approximately 15%. These findings are exceeding the adaption factor for inflammatory processes in infants the WHO uses in a current guideline[26]. We assume that during pregnancy even relatively small additional shifts in fetal iron homeostasis, especially in ferritin levels, may induce sex-specific neurodevelopmental effects[27]. Our observations regarding the link between PS, fetal iron homeostasis and the postnatal neurodevelopmental trajectories warrant further investigations, because this condition may be corrected therapeutically via targeted prenatal and/or postnatal iron supplementation[2].

The PS effect transmitted by maternal cortisol on the fetal neurodevelopment may depend on the time course of exposure[28,29]. Hypothetically, taking our explanation further antepartum, i.e., to ~3.5 weeks earlier at the time of taECG recording, we speculate that PS-induced differences in hepcidin at that time may lead to the reported changes in iron parameters that could still be detected in the cord blood[1]. Our exploratory findings of higher FSI within certain ranges of at-birth iron biomarkers support this notion. The absence of group differences of cord blood iron parameters including the whole cohort may reflect adaptions (more pronounced in females) that occur as pregnancy progresses[20].

## The role of the immune system

Our data in leukocytes showed no evidence of increased inflammatory processes in SG neonates (Table S1). Nevertheless, acute inflammatory processes, a common phenomenon during delivery, may have had an effect on our cord blood findings transmitted by other cellular messengers such as the cytokine IL-6 (Fig 1). In general, inflammation upregulates the acute phase protein ferritin influencing its role as a biomarker of the iron storage[30]. Inflammation also upregulates hepcidin levels leading to an intestinal sequestration of iron[14]. Cord blood interleukin levels were increased in chronically stressed mother's infants[31]. Taken together, the effects of PS can be mediated by inflammatory processes and this link should be investigated further in future studies including a broader characterization of the maternal and neonatal inflammatory profiles[32].

### FSI as a potential biomarker of PS in late gestation

The present findings confirm that FSI is increased in PS during the third trimester of pregnancy[10]. Because the FSI showed poor association with the measured iron biomarkers, we assume that PS influences fHR and mHR coupling by different pathways. Moreover, in our DAG framework it is conceivable that FSI may serve as an indicator of subsequent altered neurodevelopmental trajectories, even in the absence of biochemical PS correlates such as alterations in iron homeostasis[33,34].

### ML-based predictions of PS

With our ML approach, we mimicked a real-life scenario to identify mother–fetus dyads affected by PS. Our results are consistent with findings in other clinical settings where electronic medical record mining identified patients at risk even without additional biophysical assessments, such as ECG[35]. Notably, adding biophysical characteristics improves ML model performance, thus emphasizing the potential of antepartum mother–child monitoring using taECG to improve the early detection of health abnormalities such as PS.

## Strengths and Limitations

Strengths of the FELICITy study are the prospective design preventing recall bias and the definition of criteria for a matching system to exclude possible confounders. Additionally, eventual confounding factors such as the intake of iron supplement and ethnic group showed no group differences (Fig S1). This is the first prospective longitudinal study starting in utero aiming to assess PS and fetal biomarkers. Additionally, it is the first study to use causal inference and machine learning approaches to investigate sex-dependent influence of PS on the fetal iron homeostasis. There are certain limitations. Our inclusion criteria prevented us from enrolling non-German-speaking patients. This may have biased how the PS effects are represented in the multicultural Munich population. Also, we used a matching system that could not include every screened CG patient. Due to the uncertainties of a human study, several subject numbers for different sub-analyses were lower. Furthermore, we focused on measuring PS in the third trimester which necessarily neglected earlier stages of pregnancy and a possible temporal dynamic of PS over the entire course of pregnancy.

We chose not to include other potential effect modifiers on the causal pathway of the DAG such as inflammatory processes as they are difficult to define quantitatively and were not the focus of this study. However, future studies could further refine estimates of *PS* → *Iron Biomarker* average exposure effect by adjusting for these covariates.

This study did not differentiate between arterial or venous origin of the analyzed cord blood samples. To our knowledge this issue has not been addressed in literature so far. In general, the placental iron transfer and the assessment of the fetal iron status using cord blood parameters are poorly understood[36,37]. As of the date of the manuscript's submission no commonly used normal ranges of cord blood iron parameters exist. The established ranges start with the child's birth[26] but are not applicable to cord blood ranges since in cord blood usually, iron parameters are higher[38]. These issues warrant further research to identify potential biasing effects on cord blood analysis.

# Conclusions

We show that during third trimester PS exerts a sex-dependent effect on fetal iron homeostasis and on the fetal ANS measured by FSI, an ECG-derived measure of chronic stress transfer from mother to fetus. The reported biomarkers open novel avenues of research into the association between PS and adverse neurodevelopmental outcomes. They can contribute to development of novel therapeutic intervention strategies[39,40].

We propose the following aspects of future research: First, do the changes we report represent a healthy or maladaptive response to PS? Second, will applying FSI monitoring during pregnancy permit to track a "deviating neurodevelopmental trajectory" and when exactly during gestation do these changes occur? The non-invasive fetal monitoring with FSI tracking may help answer these questions. Third, what is the significance and therapeutic opportunity of the discovery that PS can impact fetal iron homeostasis? Do female fetuses possess more successful compensatory mechanisms in response to PS than males do? The sex-specific effects of this impact warrant further investigations. Fourth, which role may iron supplementation play in this context as a corrective therapeutic option[2,41]? Our findings indicate that we will need to consider the sex when devising therapeutic strategies to compensate for the intrauterine adversity due to PS[42].

# Methods

## Ethical Considerations

The study protocol is in strict accordance with the Committee of Ethical Principles for Medical Research from Technical University of Munich (TUM) and has the approval of the "Ethikkommission der Fakultät für Medizin der TUM" (registration number 151/16S). ClinicalTrials.gov registration number is NCT03389178. Written informed consent was obtained from each subject for participation in this study after having read an informative brochure and after PS screening via questionnaires and before data collection in the third trimester.

## Procedures

### Study design and study population

A prospective matched double cohort study was performed between June 2016 and July 2019 at the Department of Obstetrics and Gynecology at "Klinikum rechts der Isar" of the TUM, Germany (Fig 4). We screened 2000 women using the validated German version of "Cohen Perceived Stress Scale-10" (PSS-10) questionnaire[43]. This test quantifies the overall chronic stress based on 10 items including anxiety, depression, abnormal fatigue, and general dissatisfaction as symptoms of a generally perceived stress[44]. By means of PSS-10 we classified the patients into either stressed group (SG) or control group (CG) using a cutoff PSS-10 score of ≥19[10].

Singleton pregnant women between 18 and 45 years of age in their third trimester (at least 28 weeks gestation) were included. Exclusion criteria were serious placental alterations (e.g., IUGR), fetal malformations, maternal severe illness during pregnancy[45], preterm birth, cord blood pH < 7.10, and maternal drug or alcohol abuse. The CG (n = 85; PSS-10 < 19) was additionally matched with SG patients (n = 79; PSS-10 ≥ 19) for parity and gestational and maternal age at study entry (Fig 4). 728 participants returned the PSS-10 questionnaire and a total of 164 pregnant women were recruited.

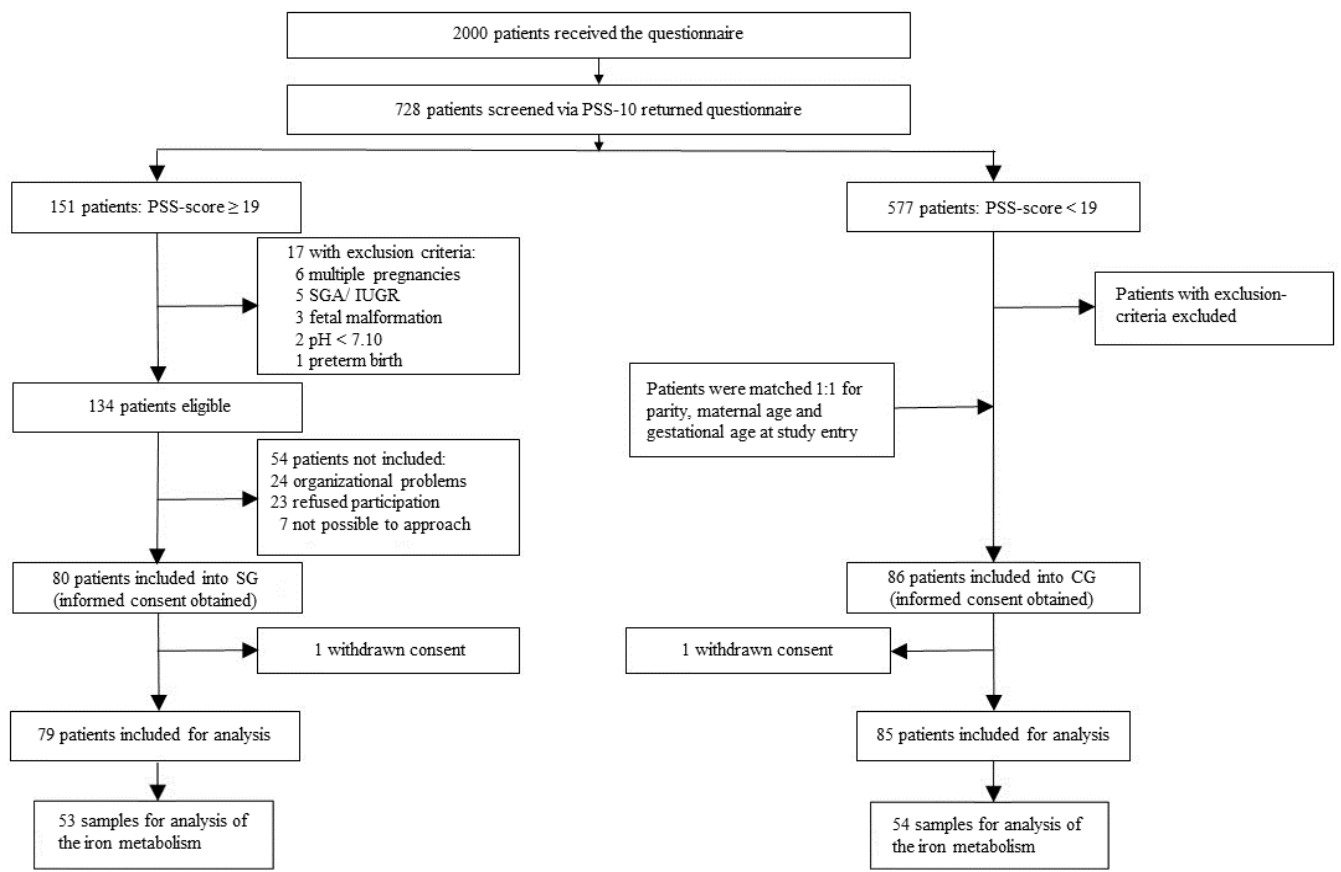

**Figure 4. Recruitment flow chart.**

**Data collection**

Participants received a sociodemographic questionnaire and a transabdominal electrocardiogram (taECG) recording at 900 Hz sampling rate of at least 40 min duration using AN24 (GE HC/Monica Health Care, Nottingham, UK). Maternal and fetal ECGs were extracted from taECG signal as described before[10]. FHR and mHR coupling was estimated using the bivariate phase-rectified signal averaging method yielding the FSI [46] which provided a measure for the fetal response to mHR decreases[10]. FSI data was evaluable from n = 139 study subjects (SG: n = 74; CG: n = 65).

During delivery, cord blood samples from a total of n = 107 patients (SG: n = 53; CG: n = 54) were extracted (Fig 4). EDTA tubes were directly analyzed to receive an adequate hemogram and serum samples were stored at –80°C until analysis[47,48]. Serum iron, transferrin, and ferritin were measured at the internal clinical laboratory of "Klinikum rechts der Isar," and hepcidin was determined using the commercial competitive "Hepcidin 25 (bioactive) HS ELISA" (DRG Instruments GmbH, Marburg, Germany).

Maternal hair samples were taken during postnatal hospitalization at the posterior vertex region of the scalp[49] for cortisol measurement using auto-analyzers[50]. Cortisol levels in 3-cm hair samples reflect chronic stress exposure of approximately 3 months prior to delivery.

After childbirth, the CG and SG perinatal outcomes were assessed. Covariates reviewed were gestational age, maternal age, gravidity, body-mass index (BMI), ethnicity, nicotine use, socioeconomic status (SES), birth weight, sex, Apgar score, and cord blood gas analysis. Additionally, clinical routine laboratory parameters were recorded including maternal hemoglobin and anemia status at the moment of hospital admission for delivery[51].

**Statistics for between-group comparison**

Continuous data were tested via Shapiro–Wilk test for normal distribution. We used t-tests for independent samples to compare SG and CG when data followed a Gaussian distribution. For non-normally distributed data, the Mann–Whitney U test was performed. Pearson's Chi-squared test compared binary coded data, and Spearman's rank-order correlation examined the relations between two variables. To assess exploratively differences in FSI in relation to iron biomarkers due to PS, we binned the value distribution of each iron biomarker automatically into five categories of equal width and compared the corresponding FSI values within each small subset. We used generalized estimating equations (GEE) to model the main effects of sex and study group and their interaction (sex*group) on iron biomarkers. All statistical tests executed were two-sided, and we assumed a significance level (α) of 0.05. IBM SPSS Statistics for Windows, version 25 (IBM Corp., Armonk, NY, USA) and Exploratory version 6.2.2 were used for modeling, statistical analysis, and visualization.

**Causal graph analysis**

We used DAGitty (www.dagitty.net)[52] to construct a directed acyclic graph (DAG) which defined the causal relationships between exposure (PS) and the two outcome measures: Iron biomarkers and a composite measure of neurodevelopment using the German adaptation of "Bayley Scales of Infant and Toddler Development – Third Edition" (Bayley Score) (Fig 3). The Bayley score has not yet been calculated for the present cohort, however we included it as an outcome measure in the causal inference framework to allow for adjustment of unobserved variables and to prevent conflict with the iron biomarker outcome. The purpose of the DAG was to visualize the structural relationships between variables and minimize bias in our statistical analysis[53] (additional Explanation in Supplement N1). The estimation of the total PS effect on both outcomes was performed in SAS version 3.8 using the "CAUSALGRAPH" function with robust error estimation.

**Machine learning (ML)**

To test the clinical utility and relative contribution of the measured parameters we classified the participants as SG or CG using ML approaches on the collected demographic, clinical, biophysical (i.e., FSI) and biochemical features (scikit-learn on Dataiku DSS 8.0.2)[54]. Binary features were expressed as categorical variables and dummy-encoded. No imputation was undertaken; rather, the missing rows were dropped. Using the conventional 80 : 20 data split for training : validation, we tested the classification performance of the following algorithms: random forest, gradient tree boosting, logistic regression, decision tree, K nearest neighbors (grid), extra trees, artificial neural network, LASSO-LARS, SVM and SGD. Threefold cross-validation with randomized grid search was used for hyperparameter optimization and fivefold cross-validation to rank the models created by each algorithm on AUROC (area under receiver operating curve). The highest-ranking model was selected and reported for each test.

# Acknowledgments

We are grateful to all female patients who participated in the FELICITy study while having to deal with the challenging last weeks of pregnancy and first years of childcare.

**Author contributions**: PZ: patient recruitment, data collection and management, data analysis, machine learning analysis and manuscript writing and editing. MF: data collection and management, data analysis, machine learning analysis and manuscript writing and editing. SML and MCA: protocol and project development, data collection and management, data analysis, and manuscript writing and editing. NW: DAG analysis, manuscript reviewing and editing. RS: manuscript reviewing and editing, CZ: patient recruitment and data collection. BF: cortisol analysis in hair samples and manuscript editing.

**Sources of support:** MCA was awarded with the August Wilhelm Scheer Professorship Program (TUM) twice for a 6-month period stay at the Klinik und Poliklinik für Frauenheilkunde, Technische Universität München, Klinikum rechts der Isar, Munich for the start-up of the FELICITy project and a Hans Fischer Senior Fellowship from IAS-TUM (Institute for Advanced Study-TUM).

**Conflict of interest statement:** MGF holds patents on fetal ECG technologies. No other disclosures were reported.

# Figure Legends

**Figure 1. Putative link of PS with iron homeostasis and maternal–fetal heart rate coupling.**

A proposed simplified model of the hepcidin–placental–ferroportin axis based on the articles of Cortes et al.[19] and Lobmaier *et al.*[10]: Hepcidin is the key regulator within this system. By binding to ferroportin, it prevents the release of iron into the bloodstream and therefore the synthesis of the transport protein transferrin and the storage protein ferritin. Hepcidin levels are influenced strongly by inflammatory processes, especially the cytokine IL-6. Another possible pathway influencing the child's stress phenotype is assumed to be realized by interference in the coupling between maternal and fetal heart rate (mHR, fHR) as seen during maternal expiration.

PSS: Perceived stress scale; FSI: Fetal stress index

**Figure 2. Sex-dependent group difference in cord blood serum ferritin levels.**

GEE model the main effects of sex and study group and their interaction (sex*group) on ferritin. GEE ferritin: group*sex **p = 0.038**

GEE: Generalized estimating equations; SG: Stressed group; CG: Control group

**Figure 3. Maternal age, socioeconomic status, and education as confounding factors within the prenatal stress trajectory.**

Directed acyclic graph analysis of the relationships between maternal and fetal antenatal, perinatal, and postnatal exposures, covariates, and outcomes.

PSS: Perceived stress scale; FSI: Fetal stress index; HR: Heart rate; BMI: Body-mass index; IVF: In-vitro-fertilization

**Figure 4. Recruitment flow chart.**

SG: Stressed group; CG: Control group; PSS: Perceived stress scale; SGA: Small for gestational age; IUGR: Intrauterine growth restriction

# Supplementary Online Content

**Table S1. Study outcome parameters**

**Table S2. Iron parameters**

**Table S3. Sex-specific linear regression of FSI and serum iron parameters**

**Table S4. Study outcome parameters for minimum adjustment sets: effect on serum ferritin and role of sex**

**Fig S1. Sex-dependent group difference in cord blood serum transferrin saturation levels**

**Fig S2. Machine learning feature importance ranking contributing to classification of stressed group and control group participants**

**Supplement N1. Additional information about causal inference analysis and causal diagrams**

**Table S1. Study outcome parameters**

| Characteristics | CG | SG | |
|---|---|---|---|
| | n=85 | n=79 | p |
| **Baseline** | | | |
| Gestational age at screening [weeks] | 34.0 (33.3–35.0) | 34.0 (32.6–34.9) | 0.304 |
| Gestational age at inclusion [weeks] | 36.7 (35.2–37.6) | 36.4 (35.3–37.4) | 0.612 |
| Age mother at study entry [years] | 33.4 (±3.7) | 32.7 (±5.1) | 0.307 |
| BMI at study entry [kg/m²] | 26.3 (24.4–28.9) | 27.8 (25.3–34.6) | **0.010** |
| BMI pregestational [kg/m²] | 21.5 (20.2–23.5) | 23.3 (20.7–27.5) | **0.013** |
| Score PSS | 9 (6–12) | 22 (20–24) | **<0.001** |
| Cortisol in maternal hair [pg/mg] | 88 (40–133) | 97 (61–165) | 0.104 |
| European/Caucasian | 78 (92) | 73 (92) | 0.879 |
| Married | 67 (80) | 55 (70) | 0.136 |
| University degree | 65 (77) | 46 (58) | **0.013** |
| Household income> 5000€/month | 49 (58) | 28 (35) | **0.004** |
| Smoking | 1 (1) | 7 (9) | **0.022** |
| Multiparity | 37 (44) | 38 (48) | 0.557 |
| Planned pregnancy | 75 (93) | 53 (67) | **0.001** |
| IVF / ICSI | 9 (11) | 2 (3) | **0.039** |
| Gestational diabetes | 2 (2) | 12 (15) | **0.003** |
| Autoimmune disease | 6 (7) | 13 (16) | 0.060 |
| Working status at screening | 3 (4) | 4 (5) | 0.502 |
| Iron supplement | 31 (36) | 35 (44) | 0.307 |
| FSI* | –0.01 ((–0.36)–0.34) | 0.38 ((–0.22)–0.75) | **0.024** |
| **Perinatal outcome** | | | |
| Gestational age at birth [weeks] | 39.9 (39.0–40.6) | 39.5 (38.6–40.6) | 0.148 |
| Birthweight [g] | 3526.9 (395.1) | 3484.0 (463.0) | 0.526 |
| Birthweight percentile [%] | 49.0 (28.3–71.8) | 55.0 (28.0–74.3) | 0.863 |
| Length [cm] | 52.9 (±2.5) | 52.8 (±2.6) | 0.919 |
| Head circumference [cm] | 35 (34–36) | 35 (34–36) | 0.412 |
| Cesarean delivery | 17 (20) | 27 (35) | **0.035** |
| Labor induction | 15 (18) | 19 (24) | 0.310 |
| Gender female | 41 (48) | 30 (38) | 0.137 |
| 5-min Apgar<7 | 3 (4) | 2 (3) | 0.691 |
| Admission to NICU | 3 (4) | 3 (4) | 0.912 |
| **Arterial plasma cord blood analysis results** | | | |
| Base Excess [mmol/L]  (n=78 CG, n=72 SG) | –5.5 (±3.3) | –5.2 (±3.0) | 0.557 |
| Lactate [mmol/L]  (n=53 CG, n=50 SG) | 4.4 (3.0–5.3) | 3.8 (3.0–4.8) | 0.317 |
| Glucose [mg/dL]  (n=57 CG, n=51 SG) | 84.0 (64.0–98.0) | 71.0 (63.5–91.5) | 0.338 |
| pH  (n=81 CG, n=77 SG) | 7.26 (±0.09) | 7.28 (±0.08) | 0.203 |
| PO2 [mmHg]  (n=66 CG, n=57 SG) | 21.1 (16.7–26.6) | 18.4 (13.6–23.5) | 0.102 |
| PCO2 [mmHg]  (n=69 CG, n=64 SG) | 50.8 (±10.2) | 49.4 (±9.2) | 0.382 |
| Leukocytes [G/L]  (n=53 CG, n=49 SG) | 14.6 (11.9–17.4) | 13.3 (10.2–17.6) | 0.291 |
| Neutrophils [%]  (n=53 CG, n=48 SG) | 51.0 (46.5–56.0) | 54.0 (47.0–61.0) | 0.249 |

Data are mean (SD) using t-test, median (interquartile range) using Mann-Whitney U test or n (%) using Pearson's Chi-squared test. Sample size is indicated as applicable. Differences with p-value < 0.05 are in bold.

PSS: Perceived stress scale; PDQ: Prenatal distress questionnaire; BMI: Body-mass index; NICU: Neonatal intensive care unit; ICSI: Intracytoplasmic sperm injection; IVF: In-vitro-fertilization

*missing values for 11 CG and 14 SG

**Table S2. Iron parameters**

| Characteristics | CG | SG | |
|---|---|---|---|
| | n=54 | n=53 | p |
| Cord blood serum iron [μg/dL] | 151.5 (±37.3) | 141.4 (±38.5) | 0.172 |
| Cord blood serum transferrin [mg/dL] | 176.3 (162.2–205.9) | 186.6 (165.8–217.0) | 0.348 |
| Cord blood serum transferrin saturation [%] | 59.5 (±17.6) | 54.8 (±19.3) | 0.189 |
| Cord blood serum ferritin [μg/L]* | 242.4 (140.6–329.6) | 176.0 (106.4–267.0) | 0.134 |
| Cord blood serum hepcidin [ng/dL] | 23.6 (13.4–39.24) | 18.9 (9.2–36.9) | 0.184 |
| Cord blood plasma hemoglobin [mg/dL]** | 15.6 (±1.6) | 15.7 (±1.6) | 0.832 |
| Cord blood plasma MCV [fL]** | 104 (101–106) | 104 (100–107) | 0.734 |
| Cord blood plasma MCH [pg]** | 35 (34–35) | 35 (34–35) | 0.605 |
| | n=74 | n=71 | p |
| Maternal prenatal plasma hemoglobin [mg/dL] | 12.3 (1.0) | 12.2 (1.1) | 0.376 |
| Maternal prenatal plasma MCV [fL] | 87 (84–90) | 88 (83–90) | 0.766 |
| Maternal prenatal plasma MCH [pg] | 30 (28–31) | 30 (28–31) | 0.862 |
| Maternal prenatal anemia: Hb<11mg/dL | 8 (10.8) | 9 (12.7) | 0.537 |
| Maternal postnatal plasma hemoglobin [mg/dL]*** | 11.1 (1.5) | 10.8 (1.3) | 0.164 |
| Maternal postnatal plasma MCV [fL]*** | 87 (85–91) | 88 (85–91) | 0.669 |
| Maternal postnatal plasma MCH [pg]*** | 30 (29–31) | 30 (28–31) | 0.683 |

Data are mean (SD) using t-test, median (interquartile range) using Mann-Whitney U test or n (%) using Pearson's Chi-squared test.

*missing values for 1 SG

**missing values for 1 CG and 4 SG

***missing values for 3 SG

**Table S3. Sex-specific linear regression of FSI and serum iron parameters**

| Characteristics | Male newborns (n=58) | | Female newborns (n=49) | |
| --- | --- | --- | --- | --- |
| | R² | p | R² | p |
| Iron [µg/dL] | 0.001 | 0.78 | <0.001 | 0.96 |
| Transferrin saturation [%] | 0.002 | 0.89 | <0.001 | 0.94 |
| Ferritin [µg/L]* | 0.003 | 0.96 | 0.048 | 0.13 |
| Hepcidin [ng/dL] | 0.028 | 0.23 | 0.004 | 0.84 |

Correlations were performed using Spearman's rank correlation.

*missing values for 1 SG with male sex

**Table S4. Study outcome parameters for minimum adjustment sets: effect on serum ferritin and role of sex**

| Parameter | Exposure Group | Estimate of Ferritin (μg/L) | Robust Std. Error | 95% Confidence Limits | | Z | Pr > \|Z\| |
|---|---|---|---|---|---|---|---|
| **Adjustment for "Household income > 5000€/month" and "Maternal Age."** | | | | | | | |
| POM | CG | 245.12 | 21.00 | 209.28 | 280.95 | 13.41 | <.0001 |
| POM | SG | 212.08 | 21.00 | 170.92 | 253.24 | 10.10 | <.0001 |
| Average Exposure Effect | | –33.04 | 27.93 | –87.79 | 21.71 | –1.18 | 0.2369 |
| **Adjustment for "University Degree" and "Maternal Age."** | | | | | | | |
| POM | CG | 246.76 | 17.08 | 213.28 | 280.25 | 14.44 | <.0001 |
| POM | SG | 208.70 | 16.85 | 175.68 | 241.72 | 12.39 | <.0001 |
| Average Exposure Effect | | –38.06 | 21.35 | –79.91 | 3.78 | –1.78 | **0.0746** |
| **Adjustment for "Household income > 5000€/month", "Maternal Age" and "Fetus Sex."** | | | | | | | |
| POM | CG | 245.15 | 19.45 | 207.03 | 283.28 | 12.60 | <.0001 |
| POM | SG | 212.48 | 16.44 | 180.27 | 244.69 | 12.93 | <.0001 |
| Average Exposure Effect | | –32.68 | 24.31 | –80.32 | 14.96 | –1.34 | 0.1788 |
| **Adjustment for "Household income > 5000€/month", "University Education" and "Fetus Sex."** | | | | | | | |
| POM | CG | 248.07 | 20.48 | 207.93 | 288.21 | 12.11 | <.0001 |
| POM | SG | 213.01 | 20.61 | 172.60 | 253.41 | 10.33 | <.0001 |
| Average Exposure Effect | | –35.06 | 26.73 | –87.44 | 17.33 | –1.31 | 0.1896 |

Differences in Average Exposure Effect with p-value < 0.1 are in bold.

POM: Potential Outcome Model

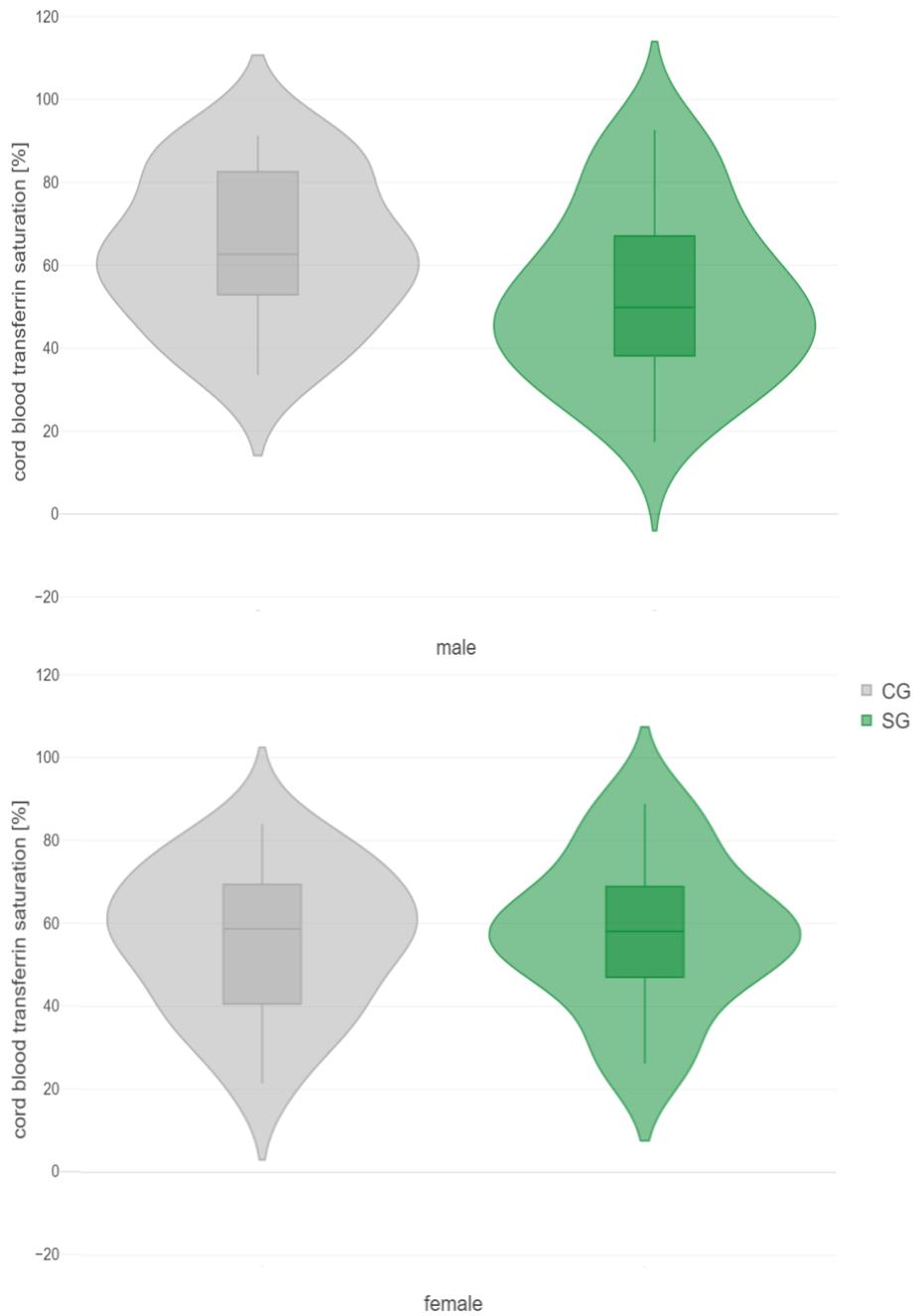

**Fig S1. Sex-dependent group difference in cord blood serum transferrin sat. levels**
GEE model the main effects of sex and study group and their interaction (sex*group) on transferrin sat. levels.
GEE transferrin: group*sex p = 0.070

SG: Stressed group; CG: Control group; GEE: Generalized estimating equations

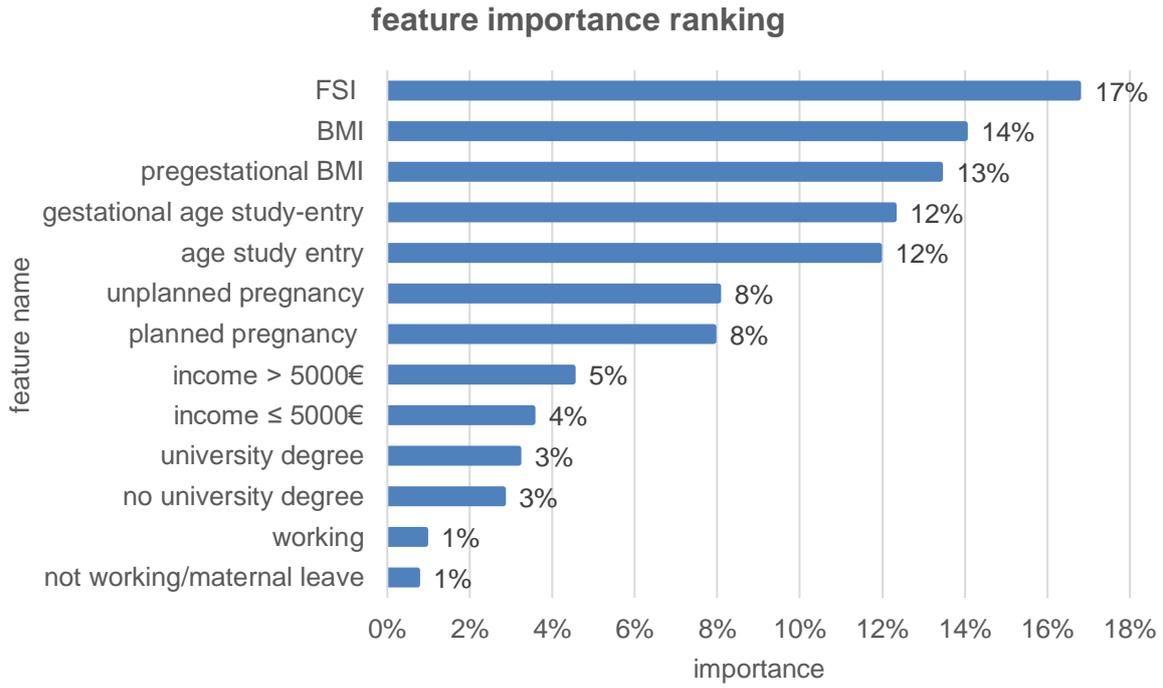

**Fig S2. Machine learning feature importance ranking contributing to classification of stressed group and control group participants**

FSI: Fetal stress index; BMI: Body-mass index

**Supplement N1. Additional information about causal inference analysis and causal diagrams**

The causal diagram shown in Fig. 3 was built as a conceptual model to demonstrate the authors' assumptions about the factors influencing the neonates' health outcome. As work by Greenland et al.[1] has demonstrated, if these diagrams are constructed according to certain rules they can provide a rigorous way to address confounding. A correctly drawn causal diagram can be used to determine whether controlling for a certain combination of variation would be sufficient to remove confounding from the exposure-outcome association, or to identify variables that should not be controlled or that need not be controlled.

As this cohort study examined multiple maternal exposures, determining a causal diagram was necessary to aid in communication with our audience *and* to ensure that we controlled for any confounding, especially as confounding can also depend on what other variables have already been controlled for.

In terms of this study, we made use of the online software "dagitty", a browser-based environment for creating, editing, and analyzing causal diagrams that makes use of the aforementioned causal diagram rules defined by Greenwood et al.

- An arrow denotes a direct causal effect: that is X→ Y implies that with all other variables held constant, changing X would change Y.
- While traditional approaches focus on individual variables, causal diagram theory focuses on how variables relate to causal paths and which set of control variables are sufficient to block those paths. In the case of Figure 3, pink arrows represent a direct causal effect along a biased path, while black arrows represent a direct causal effect along a non-biased path.
- The minimal adjustable set discussed in the manuscript is the minimum number of variables to 'close' all biasing paths (turn all pink lines to black) shown in the latter half of the new Fig. 3.

1       Greenland, S., Pearl, J. & Robins, J. M. Causal diagrams for epidemiologic research. Epidemiology 10, 37-48 (1999).